\documentclass[aps, prl, amsmath, floats, floatfix, twocolumn,
superscriptaddress, nofootinbib]{revtex4-1}

\usepackage{graphicx}
\usepackage{color}
\usepackage{soul}
\usepackage{url}
\usepackage{bm}  

\usepackage{times}
\usepackage{dcolumn}
\usepackage{bm}
\usepackage{epsf}
\usepackage{amssymb}
\usepackage{float}
\usepackage[colorlinks=true,allcolors=blue]{hyperref}
\usepackage[normalem]{ulem}
\usepackage{hyperref}

\newcommand{\beq}{\begin{equation}}
\newcommand{\eeq}{\end{equation}}
\newcommand{\beqn}{\begin{eqnarray}}
\newcommand{\eeqn}{\end{eqnarray}}

\usepackage{color}

\newcommand{\Bham}{\affiliation{School of Physics and Astronomy and Institute for Gravitational Wave Astronomy, University of Birmingham, Edgbaston, Birmingham, B15 9TT, United Kingdom
}}%
\newcommand{\UNH}{\affiliation {Department of Physics, University of New Hampshire, 9 Library Way, Durham NH 03824, USA}}
 
\newcommand{\RU}{\affiliation{Department of Astrophysics/IMAPP, Radboud University Nijmegen, P.O. Box 9010, 6500 GL Nijmegen, The Netherlands}}
\newcommand{\GRAPPA}{\affiliation{GRAPPA, Anton Pannekoek Institute for Astronomy and Institute of High-Energy Physics, University of Amsterdam, Science Park 904, 1098 XH Amsterdam, The Netherlands}}
\newcommand{\DeltaITP}{\affiliation{Delta Institute for Theoretical Physics, Science Park 904, 1090 GL Amsterdam, The Netherlands}}
\newcommand{\Nikhef}{\affiliation{Nikhef, Science Park 105, 1098 XG Amsterdam, The Netherlands}}
\newcommand{\princeton}{\affiliation{Department of Astrophysical Sciences, Princeton University, Princeton, NJ 08544, USA}}
\newcommand{\caltech}{\affiliation{Astronomy Department, California Institute of Technology, Pasadena, CA 91125}}
\newcommand{\WSU}{\affiliation{Department of Physics \& Astronomy,
	Washington State University, Pullman, Washington 99164, USA}}
\newcommand{\CITA}{\affiliation{Canadian Institute for Theoretical 
    Astrophysics, University of Toronto, Toronto, Ontario M5S 3H8, Canada}}
\newcommand{\TAPIR}{\affiliation{TAPIR, Walter Burke Institute for Theoretical Physics, MC 350-17,
    California Institute of Technology, Pasadena, California 91125, USA}}
\newcommand{\AEI}{\affiliation{Max-Planck-Institut f\"ur Gravitationsphysik (Albert-Einstein-Institut), D-14476 Potsdam, Germany}}
\newcommand{\Cornell}{\affiliation{Cornell Center for Astrophysics and Planetary Science, Cornell University, Ithaca, New York, 14853, USA}}
\newcommand{\UVirginia}{\affiliation{Department of Physics, University of Virginia, P.O.~Box 400714, Charlottesville, VA 22904-4714, USA}}

\begin{document}
\title{
Distinguishing the nature of comparable-mass neutron star binary systems with multimessenger observations: GW170817 case study
}
\author{Tanja Hinderer}\GRAPPA\DeltaITP\RU
\author{Samaya Nissanke}\GRAPPA \Nikhef\RU
\author{Francois Foucart}\UNH
\author{Kenta Hotokezaka}\princeton
\author{Trevor Vincent}\CITA
\author{Mansi Kasliwal}\caltech
\author{Patricia Schmidt}\RU\Bham
\author{Andrew R. Williamson}\GRAPPA 
\author{David A. Nichols}\RU \GRAPPA \UVirginia
\author{Matthew D. Duez}\WSU
\author{Lawrence E. Kidder}\Cornell
\author{Harald P. Pfeiffer}\AEI
\author{Mark A. Scheel}\TAPIR

\begin{abstract}
The discovery of GW170817 with gravitational waves (GWs) and electromagnetic (EM) radiation is prompting new questions in strong-gravity astrophysics. Importantly, it remains unknown whether the progenitor of the merger comprised two neutron stars (NSs), or a NS and a black hole (BH). Using new numerical-relativity simulations and incorporating modeling uncertainties we produce novel GW and EM observables for NS-BH mergers with similar masses. A joint analysis of GW and EM measurements reveals that if GW170817 is a NS-BH merger, $\lesssim 40\%$ of the binary parameters consistent with the GW data are compatible with EM observations. 
\end{abstract}
\maketitle

\begin{figure}
\includegraphics[width=0.5\textwidth]{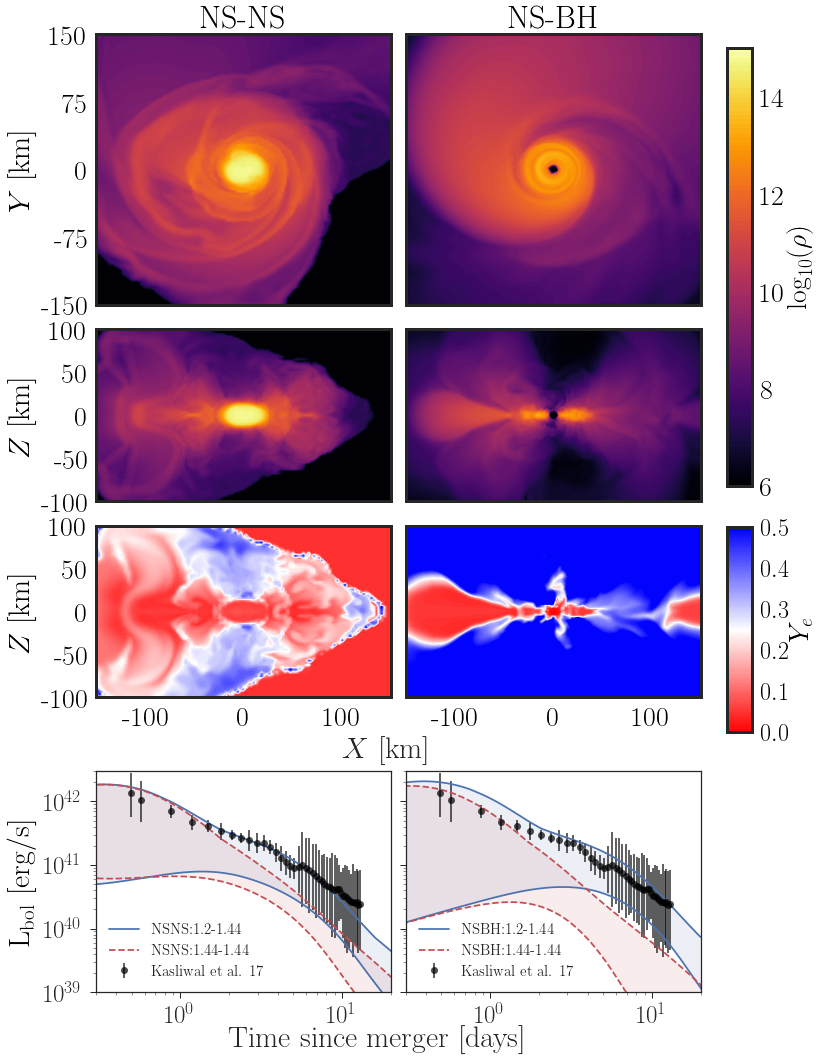}
\caption{\textit{One-to-one comparison of NS-NS and NS-BH} with $Q=1.2$ and the DD2 EoS. \textit{Upper panels}: Matter density (cgs units) and composition (electron fraction $Y_e$), $3$ms after merger for our NS-NS (left) and NS-BH (right) simulations. For NS-BH, low-density, high-$Y_e$ polar regions are not resolved numerically.
\textit{Lower panels:} Kilonova bolometric lightcurves (blue), including results for our $Q=1$ simulations (red), and observational data for GW170817 from \cite{2017Sci...358.1559K}. Shaded regions indicate the large uncertainties in the modelling. We assume a total ejecta mass of $10$--$50\%$$M_{\rm rem}$ measured in the simulations and the dynamical ejecta, 
and a $\sim 0$ -- $90\%$ fraction of the blue component, to conservatively account for uncertainties in the composition of post-merger outflows (see text).
}
\label{fig:snapshots}
\end{figure}

\textbf{\textit{Introduction}}. The recent gravitational-wave (GW) and electromagnetic (EM) measurements of GW170817~\cite{TheLIGOScientific:2017qsa, GBM:2017lvd, TheVirgo:2014hva,TheLIGOScientific:2014jea}, a neutron-star (NS) binary merger, have enabled critical insights into gravity, high-energy astrophysics, nuclear physics, and cosmology. Notably however, measurements so far have not conclusively shown that the progenitor binary comprised two NSs. From only GW observations, the individual objects' masses are consistent with current estimates of NS masses~\cite{Abbott:2018wiz}. Furthermore, under the restrictive assumption of small spins, signatures from tidal effects suggest that (at least one of) the compact objects had finite size~\cite{Abbott:2018wiz,De:2018uhw,2018arXiv180511581T}. From EM measurements alone, the discovery of a kilonova, an optical-infrared transient powered by rapid neutron-capture nucleosynthesis (e.g., \cite{1998ApJ...507L..59L,1976ApJ...210..549L,1999A&A...341..499R, 2005astro.ph.10256K,2010MNRAS.406.2650M,1999A&A...341..499R,Metzger2017}), indicates that the merger involved at least one NS ~\cite{2017Sci...358.1559K,2017Sci...358.1556C,2017ApJ...848L..19C,2017Sci...358.1556C,2017ApJ...848L..17C,2017Natur.551...80K,2017Sci...358.1583K,2017ApJ...848L..32M,2017ApJ...848L..18N,2017Natur.551...67P,2017Natur.551...75S,2017ApJ...848L..16S,2017ApJ...848L..27T,2017Sci...358.1565E}. Thus, an important open question is whether the progenitor binary was a NS-NS or a NS with an exotic compact object or black hole (BH) companion of comparable mass. A major limitation in answering the latter question has been the absence of predicted GW and EM observables for similar mass NS-BH systems. While such low-mass BHs are not expected from standard astrophysical channels, they could in principle form from primordial fluctuations in the early Universe~\cite{GarciaBellido:1996qt}; alternatively, they could be exotic objects (see, e.g.,~\cite{Barack:2018yly}). 

To address this question, this paper presents the first direct comparison between the GW and EM signatures of NS-NS and NS-BH mergers with identical mass ratios (see~\cite{Yang:2017gfb} for an initial exploration). First, using new numerical relativity (NR) simulations of nonspinning NS-NS and NS-BH mergers with an identical composition-dependent NS equation-of-state (EoS) as our benchmark, we provide GW and EM observables (GW phase evolution and EM kilonova bolometric lightcurves) for mergers with mass ratios $Q$ of $1$ and $1.2$.  Incorporating the large uncertainties in modeling as well as in the EoS of NS matter, we show that current {\emph{EM-only}} observations of GW170817 rule out an equal-mass nonspinning NS-BH merger for most realistic EoSs. We cannot, however, rule out a NS-BH merger with $Q=1.2$. Second, we use the model for the remnant mass of NS-BH mergers of Ref.~\cite{Foucart:2018rjc}, which is valid for a wide range of EoSs, mass ratios, and BH spins, to develop a general method for jointly interpreting GW and EM measurements. Third, we demonstrate that for GW170817 our {\emph{joint}} analysis leads to significantly improved constraints on the nature of the progenitor and enables us to compute, for the first time, the posterior probability distribution of NS radii and mass ratio compatible with these constraints. Our methods are orthogonal to studies that assume a NS-NS progenitor and focus on the nature of the remnant~\cite{Margalit:2017dij,Bauswein:2017,Radice:2018,Most:2018,Rezzolla:2018,Coughlin:2018miv}. For NS-NS mergers this may be either a stable or metastable NS or a BH surrounded by an accretion disk, while for NS-BH binaries can only be a BH.

\textbf{\textit{Numerical-relativity simulations.}} We analyze four new NR simulations of NS-NS and NS-BH mergers with masses $1.2M_\odot+1.44M_\odot$ and $1.44M_\odot+1.44M_\odot$, with the BH having the larger mass for NS-BH, and the tabulated composition- and temperature-dependent `DD2' EoS~\cite{Hempel:2011mk} 
for the NS matter, giving a radius $R=13$km for a $1.4M_\odot$ star. All systems are nonspinning and have low eccentricity  ($e\lesssim 10^{-3}$). Simulations are performed using the general-relativistic radiation hydrodynamics code SpEC~\cite{SpECwebsite,Duez:2008rb,Foucart:2013a}, with a two-moment approximate neutrino transport algorithm~\cite{FoucartM1:2015,Foucart:2016rxm}.  For the $Q=1.2$ systems we extract the GWs, and for all simulations we measure the mass, composition, and velocity of the matter outflows during the merger and $M_{\mathrm{rem}}$, the post-merger remnant mass excluding the final compact object. Figure~\ref{fig:snapshots} (top panels) shows the merger outcomes: matter surrounding a hypermassive NS (BH) for the NS-NS (NS-BH) systems respectively. For $Q=1$ ($1.2$) we measure $M_{\mathrm{rem}}\sim 0.08 \, (0.15)M_\odot$ for NS-NS and $M_{\mathrm{rem}}\sim 0.03 (0.12) M_\odot$ for the NS-BH binaries. In all simulations, a small amount of cold, neutron-rich material is dynamically ejected in the equatorial plane by the merger: $0.002M_\odot$ ($0.004M_\odot$) for NS-NS, and $<0.001M_\odot$ for NS-BH binaries. Less neutron-rich polar ejecta is observed, but in the absence of magnetic fields its mass is negligible (and not resolved in the simulations); see \cite{Foucartetalinprep}. Note that none of our simulations produce a relativistic jet, e.g., as observed for GW170817~\cite{Mooley:2018dlz,2018arXiv180800469G}, which is unsurprising as our simulations do not include any MHD effects (see~\cite{Paschalidis2014} for incipient jets in a NS-BH simulation).

\textbf{\textit{Tidal effects in the GWs.}}
For binaries comprising objects of a few solar masses with similar signal-to-noise ratios as GW170817, current GW detectors are sensitive only to the GWs from their inspiral~\cite{Abbott:2018wiz}. In contrast to vacuum BH-BH mergers, an important GW signature of NS matter is due to tidal effects associated with the objects' deformations. The dominant effect is characterized by the EoS-dependent tidal deformability~\cite{Flanagan:2007ix} $\lambda=(2/3)k_2 R^5/G$, where $G$ is Newton's constant, $k_2$ is the Love number and $R$ the radius.

\begin{figure}
\includegraphics[width=\columnwidth]{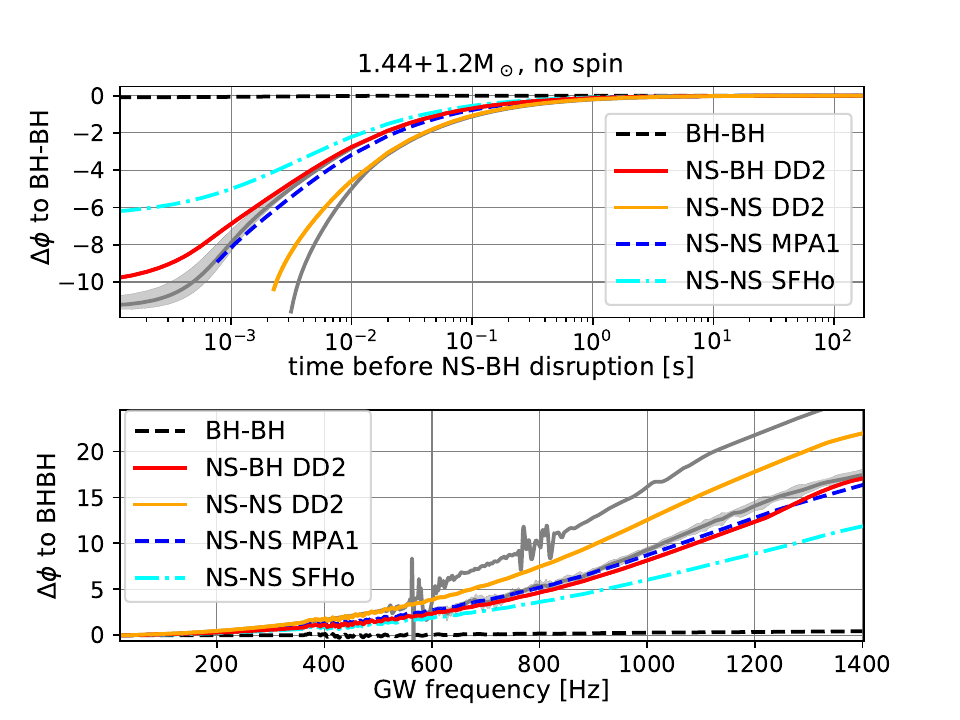}
\caption{\textit{Tidal effects during an inspiral} in the GW phase when compared to a BH-BH as a function of time (top) and GW frequency (bottom) for a $1.2M_\odot-1.44M_\odot$ system. Shown are our new NR results (grey curves) and numerical uncertainties when available (shaded regions), and predictions from the model SEOBNRv4T (curves with legends). Tidal effects accelerate the phase accumulation, hence the different signs when comparing to a BH-BH at the same time or frequency.}
\label{fig:EOB}
\end{figure}

Measurements of GW source parameters are very sensitive to the GW phase evolution (e.g.,~\cite{Cutler:1992tc,Cutler:1994ys,Veitch:2014wba}). Figure~\ref{fig:EOB} illustrates the impact of tidal effects on the GW phasing over an inspiral (from $20$Hz, where the waveforms were aligned over a $10$Hz window, up to peak GW amplitude) for a $1.44M_\odot+1.2M_\odot$ binary. Grey curves correspond to our new NR simulations, where the shaded region indicates the uncertainty due to finite resolution; the numerical errors are unimportant to our analysis below as Fig.~\ref{fig:EOB} serves merely to illustrate degeneracies between $\lambda$ (or EoS) and the type of binary. The NR data are extended to low frequencies by matching to a theoretical model (known as SEOBNRv4T~\cite{Hinderer:2016a,Steinhoff:2016rfi}), where tidal effects are described analytically and thus apply to both NS-NS and NS-BH. 
The zero-line in Figure~\ref{fig:EOB} is a BH-BH system using NR data from the SXS catalog~\cite{SXS:catalog, Blackman:2015pia} and the theoretical SEOBNRv4 model~\cite{Bohe:2016gbl,Taracchini:2013rva,Barausse:2009xi}. As seen from Fig.~\ref{fig:EOB} a NS-BH binary with the relatively stiff DD2 EoS (grey shaded region) may have similar tidal effects as a NS-NS binary with a softer EoS (smaller radius) as illustrated by dashed curves  for alternative EoS models. Together with the large statistical errors in the GW measurements, this makes distinguishing such systems difficult.

\textbf{\textit{GW170817 GW constraints}} The GW-only analysis of GW170817 allowing high spins in~\cite{Abbott:2018wiz} constrains a mass-weighted combination of tidal deformabilities $\tilde \Lambda=16/(13 M_{\rm tot}^5) [(1+12/Q)\lambda_1+(1+12Q)\lambda_2]$, where $M_{\rm tot}=m_1+m_2$ and subscripts label the objects, to be $\tilde \Lambda<630$. This bound is consistent with NS-NS, but also with BH-BH having $\tilde\Lambda=0$ and NS-BH where $\lambda_1=0$. Altogether these GW measurements can only rule out NS-BH inspirals with EoSs in extreme corners of the possible parameter space. When specializing to the more restrictive assumption of low spins, the results of ~\cite{Abbott:2018wiz,2018arXiv180511581T} are still consistent with a wide range of NS-BH binaries, including both of our simulations with the DD2 EoS~\cite{Andrewinprep}. 

\textbf{\textit{EM Kilonova observables for NS-BH and NS-NS mergers.}} For our case studies, we construct kilonova bolometric lightcurves in the ultraviolet-optical-infrared (UVOIR), arguably the most robust examples of EM observables. However, the methods presented here could be extended to any prompt emission and afterglow lightcurves associated with the short $\gamma$-ray burst (SGRB) that followed GW170817. The UVOIR lightcurve depends critically on the mass, composition and velocity of different types of matter outflow from NS-NS or NS-BH mergers~\cite{1998ApJ...507L..59L,2005astro.ph.10256K,2010MNRAS.406.2650M}, the nature of the remnant (e.g., ~\cite{Kasen:2014toa,Metzger2014}), and the inclination viewing angle to the binary (e.g., \cite{Fernandez:2016sbf}). 

We expect two types of outflows for our particular simulations: dynamical ejecta from tidal tails in the binaries' equatorial plane and winds from the remnant accretion disk. The latter strongly depend on the remnant, with an ejected mass $M_{\rm wind}\sim (0.1-0.5) M_{\rm rem}$~\cite{Just2014,Siegel2017}. Given the measured mass of the disk and dynamical ejecta, disk winds thus dominate the mass budget of the outflows. 

Based on the simulations, we compute kilonova bolometric lightcurves including conservative estimates for uncertainties in the unknown microphysics associated with the EM modelling. For simplicity, we use a two-component model with a low and high opacity component corresponding to ``blue'' and ``red''-colored parts respectively in the lightcurves (e.g., ~\cite{2017Sci...358.1583K,Metzger:2017wot}). The blue (red) components are the lanthanide-free (lanthanide-rich) ejecta with electron fraction $Y_e\gtrsim 0.25$ ($\lesssim 0.25$) \cite{2012MNRAS.426.1940K, Rosswog:2017sdn}. We solve for the evolution of the ejecta thermal energy with radiative cooling and radioactive heating \cite{Li:1998bw}. For each component, we assume that the ejecta with a total mass of $M_{\rm ej}$ and radius $r$ expand homologously with an initial density profile of $\rho \propto r^{-1}$ ($\propto r^{-5}$) for the inner (outer) part. These two parts are separated by a characteristic velocity $v_{\rm ej}$. We further assume a constant opacity with values ranging from $0.1$--$1\,{\rm cm^2/g}$  and $5$--$10\,{\rm cm^2/g}$ for the blue and red components respectively ~\cite{2013ApJ...775...18B,Tanaka:2013ana,2018ApJ...852..109T}. 

To map from the simulations to the kilonova light curves, we assume that $M_{\rm ej}$ is $M_{\rm dy}+\epsilon M_{\rm rem}$, where $M_{\rm dy}$ is the mass of the dynamical ejecta and $\epsilon=0.1$ and $0.5$ for the lower and upper bounds. The fraction of the blue component for the disk outflow ranges from $0$ (lower bound) to the value for which the slope of the bolometric light curve is consistent with the observed data (upper bound). For the dynamical ejecta we use the mass with $Y_e > 0.25$ obtained directly from the simulations. For our NS-BH simulations we obtain the upper bounds in the lower panels of Fig.~\ref{fig:snapshots} when assuming $(M_{\rm ej,red},M_{\rm ej,blue})$ of $(0.048,0.027) M_{\odot}$ and $(0.002,0.018) M_{\odot}$ for $Q = 1.2, 1$ respectively.  The lower bounds assume $(M_{\rm ej,red},M_{\rm ej,blue})=(0.015, 0)M_{\odot}$ and $(0.002, 0)M_{\odot}$
for $Q = 1.2, 1$ respectively. Correspondingly, for our NS-NS simulations, the upper bounds in Fig.~\ref{fig:snapshots} assume $(M_{\rm ej,red},M_{\rm ej,blue})=(0.032,0.02) M_{\odot}$ and $(0.006,0.02) M_{\odot}$, while the lower bounds correspond to $(M_{\rm ej,red},M_{\rm ej,blue})=(0.12,10^{-4})M_{\odot}$ and $(0.006,2 \times 10^{-4})M_{\odot}$ for $Q = 1.2, 1$ respectively. We use the electron and $\gamma$-ray heating rates of radioactive r-process nuclei given by~\cite{HotokezakaWanajo2016} and account for the thermalization efficiencies of $\gamma$ and $\beta$ rays~\cite{Barnes:2016}. Here we neglect the contribution of $\alpha$-decay and spontaneous fission.

\begin{figure}
\includegraphics[width=0.48\textwidth]{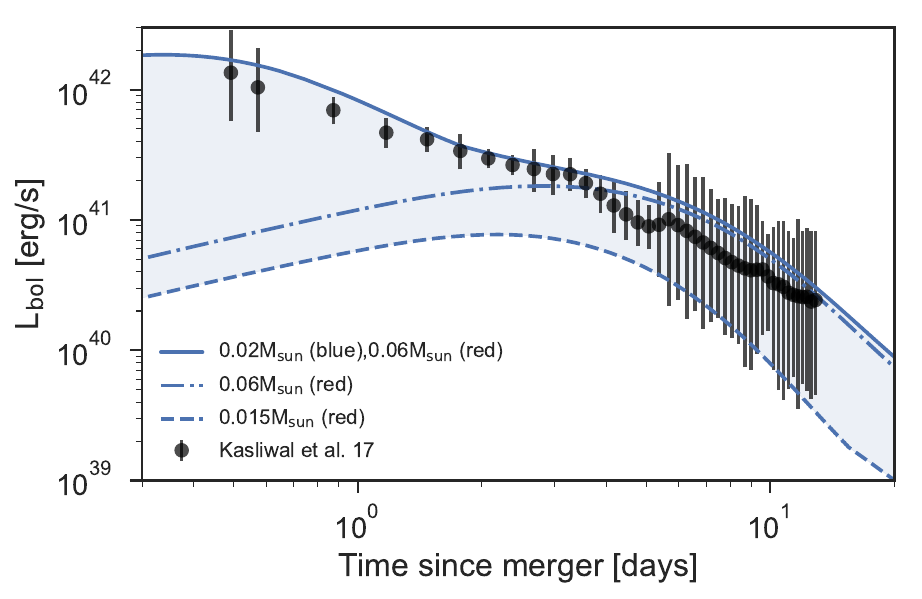}
\caption{\textit{Inferred ejecta properties required to produce the bolometric UVOIR lightcurve associated with the GW170817 progenitor}. The dotted and dashed lines show the lanthanide-rich component assuming 30$\%$ of the $(0.05 - 0.2 M_{\odot})$ remnant mass is ejected (the range in disk mass is given in our model ~\cite{Foucart:2018rjc} and the estimated ejected percentage by simulations in ~\cite{Siegel2017,Fernandez:2018}). The solid lines are the combined results from both red and blue components.}
\label{fig:EM}
\end{figure}

The bottom panels of Figure~\ref{fig:snapshots} show the kilonova bolometric lightcurves for our merger simulations together with UVOIR observations of GW170817~\cite{2017Sci...358.1559K}. 
The width of each lightcurve represents the modelling uncertainties discussed above, and uncertainties in the composition of the outflows discussed below. We find that the EM observations are \emph{inconsistent} with equal-mass NS-NS and NS-BH mergers with a DD2 EoS. They are, however, consistent with both our $Q=1.2$ NS-NS and NS-BH mergers.

\textbf{\textit{GW170817 kilonova constraints}}. Figure~\ref{fig:EM} shows the ejecta properties necessary to produce the UVOIR lightcurve associated with GW170817. The required ejecta mass can plausibly be produced by any remnant with $M_{\rm rem}\gtrsim 0.12M_\odot$ (assuming $\sim 50\%$ of the disk is unbound). Specifically, we show that the lanthanide-rich component of the lightcurve can be produced assuming 30$\%$ of $0.2 M_{\odot}$ remnant mass, given by our model ~\cite{Foucart:2018rjc} and simulations by ~\cite{Siegel2017,Fernandez:2018}, is ejected from a NS-BH merger; see \cite{Abbott:2017wuw,Coughlin:2018miv} for an alternative approach to compute photometric lightcurves for the contribution from dynamical ejecta. As discussed in~\cite{2017Sci...358.1559K,2017Sci...358.1556C,2017ApJ...848L..19C,2017ApJ...848L..17C,2017Natur.551...80K,2017Sci...358.1583K,2017ApJ...848L..32M,2017ApJ...848L..18N,2017Natur.551...67P,2017Natur.551...75S,2017ApJ...848L..16S,2017ApJ...848L..27T,2017Sci...358.1565E}, the main difficulty is to produce the $\sim 0.02M_\odot$ of fast ($v\sim 0.2$-$0.3c$), hot ejecta with a high electron fraction $Y_e\gtrsim 0.25$ required to explain the blue kilonova associated with GW170817. While none of our simulations yield such ejecta, they could be produced in the shear region between two merging NSs, though only for finely-tuned parameters~\cite{hotokezaka:13}: if the NSs' compactness is too high, the merger results in a prompt collapse to a BH preventing significant outflows, while if it is too low, the collision is insufficiently violent, yielding only a small amount of hot polar ejecta (as in our simulations). Simulations of NS-NS mergers with masses compatible with GW170817 and compactness maximizing the production of hot ejecta are necessary to determine whether such a NS-NS merger scenario can underly the blue kilonova emission associated with GW170817. 

Can the blue kilonova be produced by a NS-BH merger? While such systems do not generate polar-shocked material, they produce hot, fast ejecta through post-merger disk outflows. Outflows of the required mass, velocity, and composition are not seen in current simulations; yet these simulations suffer from important limitations. Hydrodynamics simulations of NS-BH mergers~\cite{FoucartM1:2015} show high-$Y_e$ disk winds but an insufficient amount of ejected mass; when including magnetic fields, large amounts of fast, hot ejecta have been measured~\cite{Kiuchi:2015qua}, but determining its exact mass and composition will require including neutrino transport in these simulations. Long-term magneto-hydrodynamics (MHD) evolutions of the remnant using idealized initial conditions (axisymmetric, cold, neutron-rich tori) have found fast MHD-driven outflows~\cite{Siegel2017,Fernandez:2018} but with a low $Y_e$; however, with initial conditions taken from merger simulations, 2D viscous hydrodynamics evolutions find outflows with higher $Y_e$~\cite{Fernandez:2017} than for the idealized setup. The properties of post-merger disk outflows in NS-BH systems thus remain highly uncertain. MHD effects during disk circularization and/or post-merger evolutions may still be the source of significant high-$Y_e$ outflows. 

Although these EM modelling uncertainties prevent us from setting stringent constraints on the progenitor of GW170817, we can at least rule out any binary systems that produce remnants with $M_{\rm rem} \lesssim 0.12M_\odot$. For NS-BH binaries, this critically excludes equal mass systems with $R\lesssim 13\,{\rm km}$, and compact stars ($R\lesssim 11\,{\rm km}$) at all mass ratios $Q\geq 1$, but not large stars in asymmetric-mass binaries (see below and supplementary material).

\begin{figure}
\includegraphics[width=0.45\textwidth]{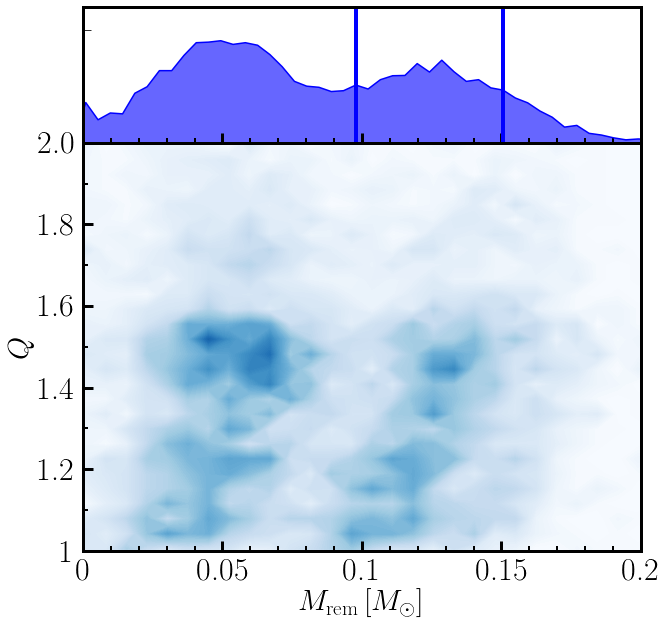}
\caption{\textit{Posterior distribution function of $Q$ and predicted $M_{\rm rem}$ for GW170817 assuming a NS-BH} merger. The top panel shows the marginalized distribution function of $M_{\rm rem}$, with the solid lines showing the $60\%$ and $90\%$ confidence intervals. The double-peaked distribution is a result of the features present in the $\tilde \Lambda$ posteriors. }
\label{fig:Mrem}
\end{figure}

\begin{figure}
\includegraphics[width=.45\textwidth]{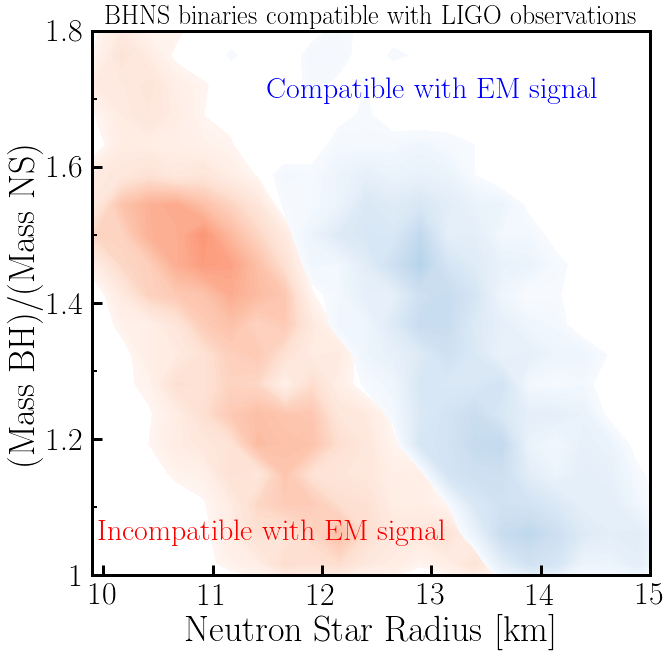}
\caption{\textit{Marginalized probability distribution of NS radii [km] and binary mass ratios for GW170817 assuming a NS-BH progenitor} and using quasi-universal relations between $R$ and $\Lambda$~\cite{Maselli:2013mva}. Current EM constraints exclude systems with $M_{\rm rem}<0.1M_\odot$ corresponding to the the red-colored part of the parameter space.}
\label{fig:RvsQ}
\end{figure}

\textbf{\textit{Joint GW and EM analysis of GW170817: a NS-BH merger?}}
When interpreting the GW and EM observations of GW170817 separately, a NS-BH binary is consistent with the measurements. Here, we show that combining GW and EM measurables yields substantially more interesting constraints on the possibility and parameters of a NS-BH progenitor. We take the posterior distributions for the effective inspiral spin $\chi_{\rm eff}$ \cite{Ajith:2009bn}, $Q$, and $\tilde \Lambda$ obtained from the GW analysis with high-spin priors from~\cite{Abbott:2018wiz}.  
Assuming a NS-BH system (zero NS spin and BH tidal deformability) we convert these parameters at fixed masses to NS deformability $\Lambda=G\lambda (Gm/c^2)^{-5}=13\tilde \Lambda (1+Q)^5/[16(1+12Q)]$ and the BH's spin parameter $\chi_{\rm BH}= (1+Q) \chi_{\rm eff}/Q$.  Using a quasi-universal relation~\cite{Maselli:2013mva,Yagi:2016bkt} we obtain the NS's compactness $C=Gm/Rc^2$ from $\Lambda$. Finally, we substitute the GW information on parameters into our model~\cite{Foucart:2018rjc} for the remnant mass $M_{\rm rem}$ given the progenitor parameters $(C,Q,\chi_{\rm BH})$. Binning these results yields the posterior distribution of $Q$ and $M_{\rm rem}$ for a NS-BH progenitor of GW170817 shown in Fig~\ref{fig:Mrem}. We find that nearly $40\%$ of the probability distribution is at $M_{\rm rem}>0.1 M_\odot$, the minimum requirement set by the EM constraints (taking into account a $
\sim 0.02M_\odot$ uncertainty in the model for $M_{\rm rem}$); see the supplemental material for the marginalized probability for a given $M_{\rm rem}$.
Figure~\ref{fig:RvsQ} shows the marginalized posterior distribution of $Q$ and $R$ for GW170817, with the region of binary parameters satisfying our conservative constraint $M_{\rm rem}>0.1M_\odot$ colored in blue. Future improved simulations of post-merger accretion disks will set both a lower and upper bound on $M_{\rm rem}$ and thus impose constraints the parameter space in Fig.~\ref{fig:RvsQ} both from bottom left and top right. Note that the region of parameter space favored by both EM and GW constraints includes equal-mass systems with large neutron stars ($Rsim 14\,{\rm km}$, also at present still consistent with nuclear physics constraints~\cite{Kolomeitsev:2016sjl}), as well as more asymmetric systems with more compact stars [e.g., $R\sim (12-13)\,{\rm km}$ for $Q\sim 1.5$].

\smallskip

\textbf{\textit{Discussion}}
We have presented the first direct comparison of NS-NS and NS-BH mergers with identical mass ratios using the results of four new NR simulations. Based on models valid over a wide range of EoSs, mass ratios, and BH spins we showed that, taking into account the large uncertainties in the EM emission and the EoS of NS matter, current GW-only or EM-only observations can rule out a NS-BH merger only in extreme corners of this parameter space. Importantly, we demonstrate a novel method for jointly analyzing GW and EM measurements to address the open question of whether one can quantitatively distinguish a NS-NS merger from a NS-BH (or exotic ultra-compact object) with comparable mass. This allows us to determine, for the first time, a quantitative result for the fraction of the NS-BH parameter space allowed by GW observations of GW170817 that is also compatible with bolometric UVOIR observations.

Our analysis is implementable for future NS binary mergers with measurable GW and EM radiation, allowing us to establish both the nature of the progenitor and remnant for single and populations of events. These methods should improve as 
simulations continue to incorporate a multitude of micro-physics, reducing the wide systematic errors in the modelling of EM measurables. In particular, our ability to predict kilonova lightcurves is severely limited by current uncertainties in the properties of the post-merger disk winds that dominate the mass budget of the outflows for near-equal mass systems. Recent progress in 3D simulations of post-merger remnants promise significant advances in modelling capabilities in the near future~\cite{Siegel2017,Fernandez:2018}. The GW measurements  will likely improve as the detectors become more sensitive, and in the more distant future may potentially observe signatures from the tidal disruption of a NS-BH system or a NS-NS postmerger signal. 

Further, our methods can readily incorporate EoS constraints from nuclear- and astrophysics (e.g., the PREX-II experiment~\cite{Fattoyev:2017jql} and the NICER mission~\cite{Watts:2016uzu}), which, when imposed, will sharpen the conclusions about the progenitor by excluding parts of the NS-BH parameter space still allowed by GW and EM observations. 
 
In conclusion, while we have focused here on the GW and EM signatures for a restricted set of NS-BH mergers, our methods have broader applications, and follow-up work is ongoing. 

\smallskip

\section{Supplemental material}

\begin{figure}
\includegraphics[width=.5\textwidth]{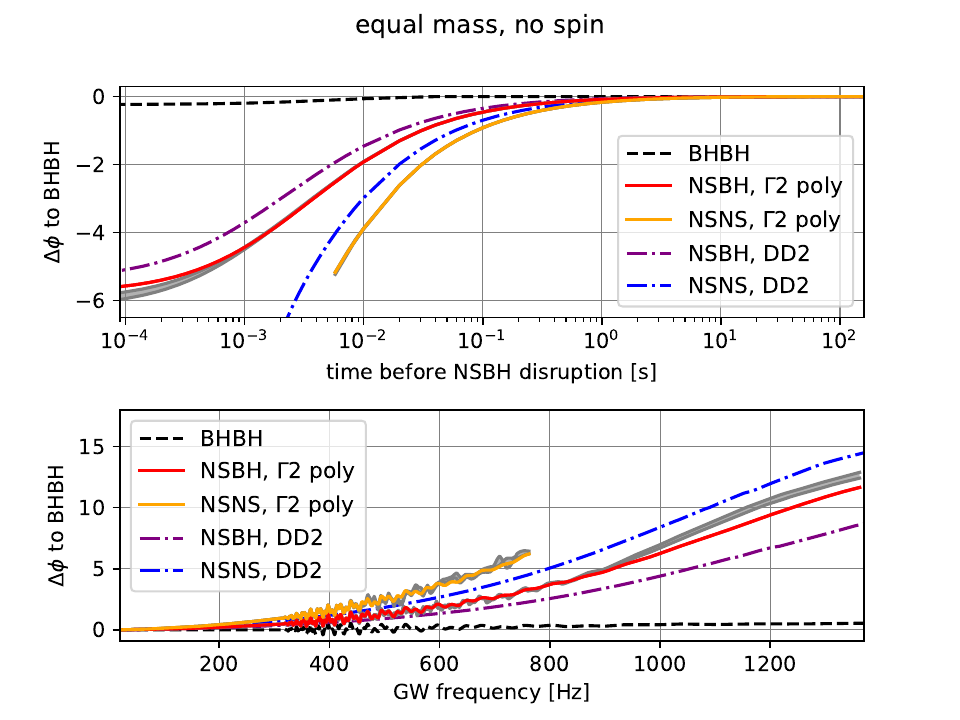}
\caption{\textit{GW phase comparisons in an equal-mass case}. All curves with legends are using the SEOBNRv4T model, shaded regions indicate the uncertainty range of NR results due to finite resolution. Note that the grey shading indicating the NR result around the orange NSNS curve is barely visible on the scale of the plot. For the DD2 cases the total mass is $M=2.88M_\odot$, while the other curves correspond to $M=2.8 M_\odot$.}
\label{fig:equalmass}
\end{figure}

\textbf{\textit{Tidal effects in the GW phasing for $Q=1$.}}
Figure ~\ref{fig:equalmass} shows a similar phasing comparison as in Fig.~\ref{fig:EOB} but for equal-mass binaries. The interesting aspect of this comparison is that for the case with a $\Gamma=2$ polytropic EoS we have accurate NR data for all types of binaries with a total mass of $M=2.8M_\odot$. For binary with the DD2 EoS discussed in the main text, $M=2.88M_\odot$. 

\begin{figure}
\includegraphics[width=0.495\textwidth]{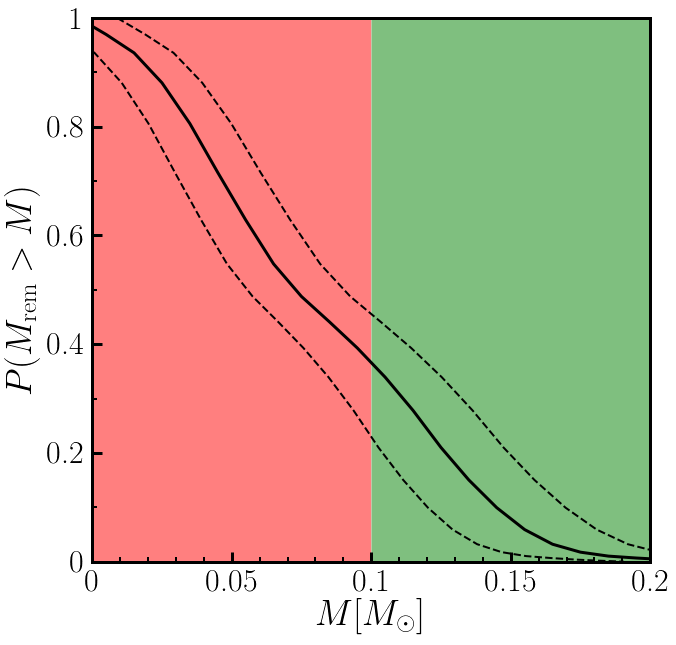}
\caption{\textit{Probability that GW170817 produces a remnant mass greater than a given value, if it is a NS-BH merger}. We show results for the model of~\cite{Foucart:2018rjc} (solid line), as well as 1-$\sigma$ errors in that formula (dashed lines).}
\label{fig:probability}
\end{figure}

\textbf{\textit{Probability of remnant mass amount for a NS-BH progenitor of GW170817.}}
The results of the combined analysis based on ~\cite{Abbott:2018wiz,Foucart:2018rjc} illustrated in Fig.\ref{fig:Mrem} can be further marginalized over the mass ratio using~\cite{Abbott:2018wiz}. We thus obtain the probability that a NSBH progenitor for GW170817 produced a given amount of remnant mass as shown in Fig.~\ref{fig:probability}. As systems with $<0.1 \, M_\odot$ of ejecta mass fail to produce the observed EM lightcurve, even under the very conservative assumptions discussed in the main text, our results show that $\lesssim 40\%$ of the parameter space allowed by the GW observations for a NS-BH progenitor of GW170817 is also consistent with the EM constraints. When more refined EM modelling becomes available in the future, Fig.~\ref{fig:probability} can be used to set tighter constraints on the possibility of a NS-BH progenitor for GW170817.  

\acknowledgements
\textbf{\textit{Acknowledgments}}
We are very grateful to the the LIGO Scientific and Virgo Collaborations for public access to their data products used in our joint GW and EM analysis of GW170817. We also thank the GROWTH collaboration for public access to their observational data products.  We thank Jacob Lange, Alexander Tchekhovskoy, and Albino Perego for useful discussions and comments, and Geert Raaijmakers for pointing out an error. TH is grateful for support from the Radboud University Excellence scheme, the DeltaITP, and NWO Projectruimte grant GWEM-NS. SMN, AW and DAN are grateful for support from NWO VIDI and TOP Grants of the Innovational Research Incentives Scheme (Vernieuwingsimpuls) financed by the Netherlands Organization for Scientific Research (NWO). FF gratefully acknowledges support from NASA through grant 80NSSC18K0565. KH is supported by the Lyman Spitzer Jr. Fellowship at Department of Astrophysical Sciences, Princeton University. TV \& HP gratefully acknowledge support by the NSERC of Canada, and the Canada Research Chairs Program. PS acknowledges NWO Veni grant no. 680-47-460. 
M.D. acknowledges support through NSF Grant PHY-1806207. L.K. acknowledges support from NSF grant PHY-1606654,
and M.S. from NSF Grants PHY-1708212, PHY-1708213, and PHY-1404569.
L.K. and M.S. also thank the Sherman Fairchild Foundation for their support. Part of this work was supported by the GROWTH (Global Relay of Observatories Watching Transients Happen) project funded by the National Science Foundation under PIRE Grant No 1545949.
Computations were performed on the supercomputer Briaree from the Universite de Montreal, managed by Calcul Quebec and
Compute Canada. The operation of these supercomputers is funded by the Canada Foundation for Innovation (CFI), NanoQuebec, RMGA and the Fonds de recherche du Quebec - Nature et Technologie (FRQ-NT). Computations were also performed on the Niagara supercomputer at the SciNet HPC Consortium. SciNet is funded by: the Canada Foundation for Innovation under the auspices of Compute Canada; the Government of Ontario; Ontario Research Fund - Research Excellence; and the University of Toronto.

\bibliography{References.bib}

\end{document}